\documentstyle[twoside,fleqn,espcrc2]{article}

\newcommand{\AmS}{{\protect\the\textfont2
  A\kern-.1667em\lower.5ex\hbox{M}\kern-.125emS}}

\title{Topological susceptibility 
through the deconfining phase transition\thanks{Pisa preprint, IFUP-TH
43/96}} 

\author{B. All\'es$^{\rm a}$ 
        M. D'Elia\address{Dipartimento di Fisica and INFN,
        Piazza Torricelli 2, 56100 Pisa, Italy}\thanks{Speaker 
        at the conference.}
        and A. Di Giacomo$^{\rm a}$
        }
       
\begin{document}

\begin{abstract}
We present a measurement of the topological susceptibility 
in $SU(3)$ Yang-Mills theory through the deconfinement phase 
transition. An improved operator is used for the topological
charge density. A drop by an order of magnitude is observed
from the confined to the deconfined phase.
\end{abstract}

\maketitle

\section{INTRODUCTION}

The topological susceptibility is defined in continuum QCD  as 
\begin{equation}
 \chi \equiv \int d^4 x \langle 0 | T(Q(x) Q(0))| 0 \rangle,  
\end{equation}
where
\begin{equation}
Q(x) = {{g^2} \over {64 \pi^2}} \epsilon^{\mu \nu \rho \sigma} 
 F^a_{\mu \nu} (x) F^a_{\rho \sigma} (x).
\end{equation}
$Q(x)$ is the topological charge density and is proportional to the anomaly 
of the $U_A (1)$ current \cite{hoft1}
\begin{equation}
\partial_{\mu} j^{\mu}_5 (x) = 2 N_f Q(x). 
\label{eq:anomaly}
\end{equation}
The value of $\chi$ for quenched QCD can be  related to the masses of the 
pseudoscalar mesons, by use of $1 \over {N_c}$ expansion \cite{wit2,vene2}
\begin{equation}
  {{2 N_f} \over {f_{\pi}^2}} \chi = m_{\eta }^2 + m_{\eta '}^2 
                       - 2 m_{K}^2.
\label{eq:massform}
\end{equation}

$\chi$ has  been determined on the lattice \cite{digia1} 
and is consistent with the prediction 
of eq. (\ref{eq:massform}). This solves the so-called $U_A (1)$ problem 
\cite{wein}. 

 It is interesting to determine 
the behaviour of  $\chi$ at $T \neq 0$ , in particular across 
the deconfining phase transition, which in quenched QCD takes place at 
$T_c \sim 260$ MeV \cite{boyd}. 
A  general expectation is that the topological susceptibility
should drop at $T_c$ \cite{pisar}, since
Debye screening inhibits tunneling between states of different 
chirality and damps the density of instantons.

Previous attempts to study $\chi$ through $T_c$  by 
the field theoretical method were plagued with large statistical uncertainties above $T_c$ \cite{digia2}. Also the attempts to use the cooling technique
en\-coun\-tered difficulties at finite $T$ \cite{digia2,tep2}. The idea of the 
cooling method \cite{tep2} is to measure the topological charge on 
configurations cooled by locally minimizing the action. 
At $T = 0$ and below $T_c$ the method works well 
and instantons show up as plateaux of the topological charge versus cooling. 
Around $T_c$ instantons become unstable, the plat\-eaux disappear and the 
method becomes am\-bigu\-ous \cite{digia2}.

In the present paper we drastically reduce the statistical fluctuations 
of the field theoretical method determination, by using an improved 
topological charge operator as proposed in ref. \cite{christou}.

We redetermine $\chi$ at $T = 0$  
and study its behaviour through $T_c$. The new determination of $\chi$ at 
$T = 0$ is consistent with previous determinations \cite{digia3,np3,tepchi}.
At finite $T$ we obtain a good de\-ter\-mina\-tion of $\chi$.  
Our main result is that $\chi$ drops at $T_c$ 
by more than one order of magnitude.






\section{THE METHOD}

The field theoretical method is a straight\-for\-ward application of the basic 
rules and concepts of quantum field theory.

At first a lattice discretization  
$Q_L(x)$ of  the continuum topological charge density 
operator $Q(x)$ is defined. 
$Q_L(x)$ is not unique: in\-fin\-itely many operators can be defined 
on the lattice which have the same continuum limit.

 In going to the continuum limit a proper renormalization  
must be performed, like in any other regularization scheme. In quenched QCD 
$Q_L$ 
renormalizes multiplicatively
\cite{z1}. In formulae  
\begin{equation}
Q_L = Z(\beta ) Q a^4 + {\cal O}( a^6 ).
\end{equation}
As usual, $\beta \equiv 6/g_0^2$.
The topological sus\-cept\-ib\-il\-ity can be defined on the lattice as 
\begin{equation}
\chi_L \equiv \langle \sum_x Q_L(x) Q_L(0) \rangle.  
\end{equation}
The standard rules of renormalization then give
\begin{equation}
 \chi_L = Z(\beta)^2 a^4 \chi + M(\beta) + {\cal O}(a^6),
\label{eq:defchilat}
\end{equation}
where  $M(\beta)$ is an additive renormalization containing
mixings of $\chi_L$ to other operators with the same  quantum numbers
and lower or equal dimensions \cite{tutti}.

$Z$ and $M$ can be computed non-perturbatively on the lattice by the 
heating method \cite{np3,np1,np3bis}.  Renormalizations are induced by 
quantum fluc\-tu\-ations at the scale of the UV cutoff (lattice 
spacing scale), so they can be estimated by meas\-uring 
topological quantities on an ensemble of configurations 
where the topological content is known and short-range fluctuations 
are thermalized. In order to do this one can start with a classical
configuration on the lattice and perform few local updating steps on it 
until the short-range fluctuations are thermalized: 
by repeating this procedure 
several times an ensemble of tra\-ject\-ories can be con\-structed, all starting
from the same initial configuration. Starting from a large instanton
 and  meas\-uring $Q_L$  on the ensemble, a 
plateau at the value of $Z(\beta )$  is reached with respect 
to the number of heating step performed. 
Similarly starting from the flat configuration and meas\-uring $\chi_L$
a plateau at the value of $M(\beta )$ is reached.
To be sure that the initial topology  is not changed by the heating 
procedure, each con\-fig\-uration of the sample is checked during heating by  
performing a few cooling steps to detect its topological charge 
on a copy of it. 
If a change occurred the configuration is discarded \cite{fp}.


Once $M$ and $Z$ are known, $\chi$ can 
be extracted by a subtraction procedure following eq. (\ref{eq:defchilat}). 
Clearly $Z$ and $M$ 
depend on the operator used for $Q_L$. A standard definition for 
$Q_L$ is \cite{divec}
\begin{equation}
Q_L(x) = {{-1} \over {2^9 \pi^2}} 
\sum_{\mu\nu\rho\sigma = \pm 1}^{\pm 4} 
{\tilde{\epsilon}}_{\mu\nu\rho\sigma} \hbox{Tr} \left( 
\Pi_{\mu\nu} \Pi_{\rho\sigma} \right).
\label{def:qstand}
\end{equation}
Here ${\tilde{\epsilon}}_{\mu\nu\rho\sigma}$ is the
standard Levi-Civita tensor for positive directions while for negative
ones the relation ${\tilde{\epsilon}}_{\mu\nu\rho\sigma} =
- {\tilde{\epsilon}}_{-\mu\nu\rho\sigma}$ holds. 
$\Pi_{\mu\nu}$ is the plaquette in the $\mu - \nu$ plane at point $x$.
With this definition
$Z \simeq 0.18$ and the mixing 
$M$ is most of the measured signal in the scaling region \cite{digia3}, 
so that a large error derives from the subtraction procedure performed to 
extract $\chi$. 

In order to obtain a more precise determination of $\chi$, 
a new definition of $Q_L$ can be used for 
which $M$ is strongly reduced and $Z$ is closer to its continuum value
$Z = 1$. 

In this work we use the sequence of improved operators $Q_L^{(i)}$ 
defined in ref. \cite{christou}: the starting operator of the sequence is the standard
def\-ini\-tion of eq. (\ref{def:qstand}) and a smearing procedure is performed at 
each step of the sequence. We stop at the second smearing, where a 
considerable im\-prove\-ment is already achieved. For a full account of our 
results see ref. \cite{we}.

\section{RESULTS}

We measured the topological susceptibility at zero temperature on
a symmetric lattice $16^4$ and at finite temperature on a lattice
$32^3\times 8$. 
The simu\-lat\-ions were performed on a APE QUADRICS machine.

In figure 1 we plot the value of $(\chi)^{(1/4)}$ at zero
temperature versus $\beta$ for the three definitions 
$\chi_L^{(i)}$ ($i=0,1,2$) of the lattice
topological sus\-cept\-ib\-il\-ity.
There is good scaling: $(\chi)^{(1/4)}$ is in\-de\-pend\-ent of $\beta$, as it should. 
There is  also an excellent agreement between the three determinations. 
To fix the value of $(\chi)^{(1/4)}$ in physical units we use 
for $a(\beta)$ the two loop formula
and for $\Lambda_L$ the determination of ref. \cite{bali} $\Lambda_L=4.56(11)$
MeV.
The horizontal line is the
linear fit to the 2-smear data. It gives $(\chi)^{(1/4)} = 175(5)$
MeV and is consistent, within errors, with that of ref. 
\cite{digia3,np3,tepchi}. 
The error includes the uncertainty in $\Lambda_L$.

\begin{figure}[t]
\vspace{3.1cm}
\includegraphics{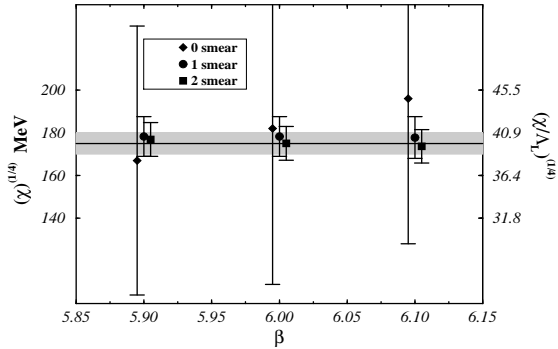} 
\null\vskip 0.3cm
\caption{$\chi$ at $T=0$.
The 
straight line is the result of the linear fit of the 2-smeared data.
The improvement from $Q^{(0)}_L$ to $Q^{(2)}_L$ is clearly visible.}
\end{figure}

In figure 2 the topological susceptibility 
$\chi \equiv (\chi^{(i)}_L - M^{(i)})/(Z^{(i)2} a^4)$
at the 
transition point is shown for the 1 and 2-smeared operator. $\chi$
drops by one 
order of magnitude from the confined to the deconfined phase. 
The results obtained with the two operators are compatible as they should.
The data for the 0-smeared operator have very large error
bars and are not shown in the figure.
The data have been plotted versus $T/T_c$ where $T_c$ is the
deconfining temperature coresponding to $\beta_c(N_{\tau}=8)=6.0609(9)$ \cite{boyd}.  
To determine $T/T_c$ we only need the ratio $a(\beta_c)/a(\beta)$ and
for that the two-loop expression is certainly a good approximation
within the small interval of $\beta$ used, where $\Lambda_L$ can be
considered as a constant.

The solid line in figure 5 corresponds to the value of $\chi$ at
zero-temperature and is consistent with the data below $T_c$
indicating that $\chi$ is practically $T$-independent in the confined
phase.
\\

We thank Graham Boyd and Enrico Meggiolaro
for useful discussions. A.D.G. ac\-know\-ledges an interesting discussion
with Heinrich ~Leutwyler.

\begin{figure}[t]
\vspace{3.1cm}
\includegraphics{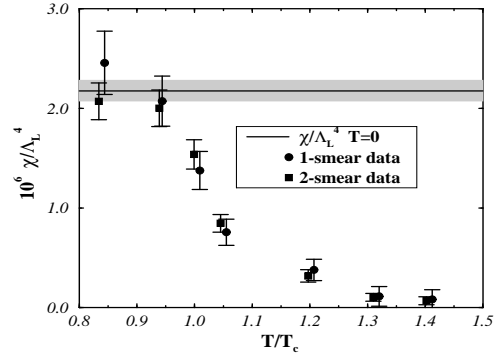} 
\null\vskip 0.3cm
\caption{$\chi/\Lambda_L^4$ versus $T/T_c$ across
the deconfining phase transition.
The horizontal band is the determination at $T=0$ of figure 1.}
\end{figure}

\end{document}